\documentstyle[12pt]{article}

\def\today{\number\day\space  \ifcase\month\or
           January\or February\or March\or April\or May\or June\or
           July\or August\or September\or October\or November\or
           December\fi
           \space \number\year}

\begin{document}

\hfill{U. of MD PP \#94-129}

\hfill{DOE/ER/40762-033}

\hfill{WM-94-104}

\hfill{hep-ph/9405344}

\hfill{May 1994}   \vskip 0.7in

\baselineskip = 18pt

\begin{center}
{\large {\bf Penguins leaving the pole: bound-state
effects in $B \rightarrow K^* \gamma$} } \vskip 0.6in

{C. E. Carlson}

{\it Physics Department, College of William and Mary

Williamsburg, VA 23187, USA }     \vskip 0.3in

{J. Milana}

{\it Department of Physics, University of Maryland

College Park, Maryland 20742, USA  }    \vskip 0.8in

\end{center}

{\narrower

Applying perturbative QCD methods recently seen to give a good
description of the two body hadronic decays of the
$B$ meson, we address the question of bound-state effects on the
decay $B\rightarrow K^* \gamma$.  Consistent with most analyses, we
demonstrate that gluonic penguins, with photonic bremsstrahlung off
a quark, change the decay rate by only a few percent.  However,
explicit off-shell $b$ quark effects normally discarded are found
to be large in amplitude, although in the standard model accidents
of phase minimize the effect on the rate.  Using an asymptotic
distribution amplitude for the
$K^*$ and just the standard model, we can obtain a branching ratio
of a few $\times 10^{-5}$, consistent with the observed rate.

}


\newpage

\section{Introduction}

This note reports on potential bound state modifications to the
analysis of $B \rightarrow K^* \gamma$, which is usually given
solely in terms the on-shell subprocess $b \rightarrow s \gamma$.

The flavor changing neutral currents involved in the decays of the
$B$ into $K^* \gamma$ do not exist to leading
order in the standard model, but can occur in second order in the
Weak interaction via emission and reabsorbtion of $W$
bosons~\cite{gw79}.  These loop diagrams are often called
``penguins,'' and their magnitude can be greatly modified by strong
interaction effects~\cite{ag93,gsw90,g89}.

There is recent further interest in these decays because
additional penguin-like contributions could come from particles
not in the minimal standard model~\cite{h93,bg93}.  Contributions to
$B$ into $K^* \gamma$ decay from loops of non-standard-model
particles (such as loops of supersymmetric particles that might be
called ``penguinos'') would be a signal of their existence.  To take
advantage of this possibility, more precise study of the decay in
the standard model needs to be undertaken.

The subprocess $b \rightarrow s \gamma$, taken as a free decay, is
usually treated as the only flavor changing contribution leading
to $B \rightarrow K^* \gamma$~\cite{bhs94}.  However, bound state
effects could seriously modify results coming from this
assumption.  Bound state effects include modifications due to the
quarks being off shell in
$b
\rightarrow s \gamma$,  contributions involving gluonic penguins
or double (photon plus gluon) penguins, and contributions from
annihilation diagrams.  The latter involve no neutral flavor
changing currents at all.

We use shall
perturbative QCD (pQCD) in our analysis (see also~\cite{g93}), a
methodology we have previously applied~\cite{cm93,cm94} to hadronic
decays and semileptonic form factors of the $B$, inspired by
Ref.~\cite{shb}.  Examples of the
Feynman diagrams we calculate are given in Fig.~\ref{diagrams}.
We require as input the effective vertices that result from the
penguin diagram analyses~\cite{gsw90,g89}.  Thereafter our
calculations are quite explicit and are detailed below.

A preview of our results is as follows:  Diagrams involving the
subprocess $b \rightarrow s \gamma$ do dominate, and keeping just
contributions from the most commonly cited effective vertex gives
close to the correct answer. However, some luck underlies the
last statement.  Since the internal heavy quark propagator can go
off shell, there is an additional, independent, effective
vertex~\cite{g89} that can contribute.  It does so with an
amplitude whose magnitude is about $2/3$ that from the usual
vertex, and only because of fortunate phase relations is the
magnitude of the sum nearly the same as for the usual vertex
alone.  Other diagrams are shown to be small, although they lead
to a someday measurable few percent difference in the decay rates
of the charged and neutral $B$ into $K^* \gamma$ decay.

A diagram we are forced to omit for now is the double penguin,
Fig~\ref{diagrams}e, where both a photon and a gluon come out of
the loop, as the effective vertex it gives is not calculated.

\section{Calculations}

We now begin to describe our calculations in more detail.  The
penguins are represented by blobs in Fig.~\ref{diagrams}, which may
be interpreted as an effective Hamiltonian,  expanded as
\begin{equation}
H_{eff} = - 4{G_F\over\sqrt 2} V_{tb} V_{ts}^*
              \sum C_i(\mu)O_i(\mu).
\end{equation}
The operators $O_i$ are listed in references \cite{gsw90} and
\cite{g89}.

Consider the photonic penguin diagrams in Fig.~\ref{diagrams}a. If
the incoming and outgoing quarks in $b \rightarrow s \gamma$ are on
shell,  there is only one relevant operator in $H_{eff}$.  Refs.
\cite{g89} contain two operators that can contribute, but using
$i{\raise 1pt \hbox{$\not$}}D q=m_q q$ they can be seen to be
equivalent. We will write the operators so that only one
contributes when the quarks are on-shell. The commonly used
operator relevant to radiative $b$ decay is then
\begin{equation}
O_7 = {e\over 16\pi^2}\ m_b \bar s \sigma^{\mu \nu} F_{\mu \nu}
    {1\over 2} \left (1+\gamma_5 \right) b.
\end{equation}
The numbering is that of Ref. \cite{gsw90}; unfortunately the
notations of \cite{gsw90} and \cite{g89} do not match.  The other
operator is
\begin{equation}
O_2^\prime = {e\over 16\pi^2}\ \bar s \sigma^{\mu \nu} F_{\mu \nu}
  {1\over 2} \left (1+\gamma_5 \right)
  (i{\raise 1pt \hbox{$\not$}}\kern-1pt D - m_b) b,
\end{equation}
where the numbering is from Ref. \cite{g89} and the prime reminds
us that we have put the on shell part into the other operator (and
that we changed the location of a factor $Q_d = -1/3$).

Let us here record the coefficients at the $W$ mass scale,
\begin{equation}
C_2^\prime(m_W) = {x\over 24(1-x)^4}\left((1-x)(18x^2-11x-1)
                   +2(3x-2)(5x-2) \ln x \right) = -0.47,
\end{equation}
\begin{equation}
C_7(m_W) = {x\over 24(1-x)^4}\left((1-x)(8x^2+5x-7)
                   +6x(3x-2) \ln x \right) = -0.19,
\end{equation}
where the numerical values are for $x \equiv m_t^2/m_W^2 = 4$, and
also record how these operators evolve down to lower scale,
\begin{equation}
C_2^\prime(\mu) = C_2^\prime(m_W)
               -{22\over 81}\left(1-\eta^{-2/\beta_0}\right)
               -{11\over 81}\left(\eta^{4/\beta_0} -1\right) ,
\end{equation}
\begin{equation}
C_7(\mu) = \eta^{-16/3\beta_0} \left[ C_7(m_W)
               -{58\over 135}\left(\eta^{10/3\beta_0} -1\right)
               -{29\over 189}\left(\eta^{28/3\beta_0} -1\right)
              \right] .
\end{equation}
Quantity $\eta$ is $\alpha_s(\mu)/\alpha_s(m_W)$ and $\beta_0$ is
$11 -(2/3)n_f$, and following standard practice we have neglected
mixing with operators that give small effect.  Both $C_7$ and
$C_2^\prime$ increase in magnitude with decreasing scale $\mu$.

We make the peaking approximation for $\phi_B$, the distribution
amplitude of the $B$ meson, wherein
\begin{equation}
\phi_B(x) = {f_B \over 2\sqrt 3} \delta (x_1-\epsilon_B).
\end{equation}
The decay constant of the $B$ is $f_B$ and $x_1$ is the
light cone momentum fraction carried by the light quark.  The mass
of the $B$ is given by $m_B = m_b + \bar\Lambda_B$ and $\epsilon_B
= \bar\Lambda_B / m_B$.  For the
$K^*$ distribution amplitude, we write
\begin{equation}
\phi_{K^*}(y) = {\sqrt 3} f_{K^*} y_1y_2 \tilde \phi_{K^*}(y).
\end{equation}
The normalization is
\begin{equation}
\int_0^1 dy_1\, \phi_{K^*}(y) = {f_{K^*} \over 2\sqrt 3}
\end{equation}
so that $\tilde \phi_{K^*}(y)$ is unity for the (super-)asymptotic
distribution amplitude. We also make the approximation that
$m_{K^*} = 0$.  The second diagram of Fig. \ref{diagrams}a is then
zero.

The spin projection operators for the initial and final hadronic
states are $\gamma_5 ({\not}p-m_B)/\sqrt 2$ and ${\not}\kern1.5pt\xi
({\not}\kern1.5pt k+(m_{K^*}))/\sqrt 2$, respectively, with $p$,
$k$, and
$q$ being the momenta of the $B$, $K^*$, and photon, and $\xi$ the
polarization vector of the $K^*$.  Angular momentum conservation
allows only transverse polarizations.

The result is
\begin{equation}
M_{\rm photonic\ penguin} = -{8G\over m_B \epsilon_B}
          \left(C_7(\mu) I - 2C_2^\prime(\mu) I^\prime\right)
          \left( p \cdot q \epsilon \cdot \xi
          + i\epsilon_{\mu\nu\alpha\beta}
          p^{\mu} q^{\nu} \epsilon^{\alpha} \xi^{\beta} \right),
\end{equation}
where $\epsilon$ is the polarization of the photon and
\begin{equation}
G = C_F {e \alpha_s\over 4\pi} {G_F\over\sqrt 2} V_{tb} V_{ts}^*
                     f_B f_{K^*}
\end{equation}
with $C_F =4/3$.  Also
\begin{eqnarray}
I &=& (1-\epsilon_B) \int_0^1 dy_1\,\tilde\phi_{K^*}(y)\,
         {(1-y_1)(1+y_1-2\epsilon_B)\over y_1-2\epsilon_B-i0^+}\\
  &=& (1-\epsilon_B) \left(  - {1\over 2}  +
      (1-2\epsilon_B)
      \left[ i\pi+\ln{1-2\epsilon_B\over{2\epsilon_B}}\right]
          \right)
\end{eqnarray}
and
\begin{equation}
I^\prime = \int_0^1 dy_1\, (1-y_1)\, \tilde\phi_{K^*}(y)
         = {1\over 2},
\end{equation}
where the integrated results are for the asymptotic distribution
amplitude.  The imaginary part come from an internal propagator
going on shell.  This is often taken as a signal that pQCD is
inapplicable.  However, if properties of the reaction overall
dictate short distance propagation only, then pQCD can still be
used~\cite{cn65+}.  This is the situation here, as discussed
in~\cite{cm93} and~\cite{islamabad}.

Numerical results will be given after we have discussed what
prove to be the subdominant contributions.  Our understanding of
why they are subdominant is helped by some order of magnitude
estimates.  The crucial expansion parameter is $1/\epsilon_B$ and
its logarithms.  Factors of $\epsilon_B$ come from the propagators,
and can also be induced, depending on circumstances, by the factor
$y_1y_2$ in the $K^*$ distribution amplitude.

In the photonic penguin diagrams, Fig. \ref{diagrams}a, the gluon
connects at the lower vertex to on-shell quarks and its propagator
gives a factor proportional to
$1/y_1\epsilon_B$, where $y_1$ and $\epsilon_B$ are the momentum
fractions of the two light quarks. Thus appears one factor of
$1/\epsilon_B$; the $y_1$ is canceled from the $K^*$ distribution
amplitude.  The $b$ quark propagator is involved in two
subprocesses: scattering from the light quark by gluon exchange
and decay into the $s$ quark plus photon.  Both are possible for
an on-shell $b$ quark, and the $b$ quark does go on-shell in
this diagram when $y_1 = 2\epsilon_B$. The $b$ quark propagator thus
contributes an imaginary pole term and a real principal value term
(or just a real term, for the operator $O_2^\prime$\,) to the
integral involving the $K^*$ distribution amplitude, and one of
them gives (roughly speaking) an $i\pi$ and the other gives a
logarithm of $1/\epsilon_B$. Now we have accounted for the
$\epsilon_B$ factors that appear in the photonic penguins,
\begin{equation}
M_{\rm photonic\ penguin}\approx
({\rm factors})
\times {1\over\epsilon_B}
\times \left (C_7(\mu)\
\times O \left (i\pi\ {\rm or}\ \ln{1\over\epsilon_B}\right)
+ \ C_2^\prime (\mu) \times O(1) \right) .
\end{equation}

The gluonic penguin graph with emission of the photon from the $b$
or $s$ quark (Fig. \ref{diagrams}b) does not allow the quark
propagator to be on shell.  For example, in the diagram with photon
emission from the $s$ quark, the internal $s$ quark decaying into
an on-shell photon and and on-shell $s$ quark must have a momentum
squared larger than the mass squared of the $s$. A similar argument
shows the $b$ propagator is never on shell in the diagram with
photon emission from the $b$ quark.  The gluon propagator is still
spacelike and still gives a $1/\epsilon_B$, but compared to the
previous case we lose the $\log(1/\epsilon_B)$ or $i\pi$ that came
from quark propagator,
 \begin{equation}
M_{\rm b\ or\ s\ bremss}\approx
({\rm similar\ factors})
\times {1\over\epsilon_B}
\times C_8(\mu)
\times O\left(1\right).
\end{equation}
As we shall see,  another significant reduction comes from the
replacement of coefficient $C_7$ by its gluonic counterpart $C_8$.

For the spectator bremsstrahlung case, Fig. \ref{diagrams}c, 	the
quark propagator cannot go on-shell.  However, the gluon propagator
can. A factor $(1/\epsilon_B)$ that came from the gluon propagator
is lost,  and replaced by factors $i\pi$ or $\log(1/\epsilon_B)$
that come from integrating the $K^*$ quarks's momentum fraction
over the gluon pole, yielding
\begin{equation}
M_{\rm spectator\ bremss}\approx
({\rm similar\ factors})
\times {(\epsilon_B)^0}
\times C_8(\mu)
\times O \left (i\pi\ {\rm or}\ \ln{1\over\epsilon_B}\right).
\end{equation}
Thus the gluonic penguin diagrams are suppressed by powers of
$\epsilon_B$ or logs thereof, as well as by the ratio $C_8/C_7$.

For the actual calculations involving the gluonic penguin we kept
just $O_8$ in the effective hamiltonian, where
\begin{equation}
O_8 = {g\over 16\pi^2}\ m_b \bar s \sigma^{\mu \nu} G_{\mu \nu}
    T_a {1\over 2} \left (1+\gamma_5 \right) b.
\end{equation}
The numbering is that of Ref. \cite{gsw90}. Other
operators are possible when the gluon or quarks are off-shell.  We
have not explicitly calculated their contributions in this case
(in part because their evolution has not been calculated),  but have
verified that the order of magnitude estimates are not upset,
i.e., they are not leading in $1/\epsilon_B$.

The results are
\begin{equation}
M_{\rm b\ or\ s\ bremss} = {8 e_d G\over m_B \epsilon_B}
         C_8(\mu) I^\prime
          \left( p \cdot q \epsilon \cdot \xi
          + i\epsilon_{\mu\nu\alpha\beta}
          p^{\mu} q^{\nu} \epsilon^{\alpha} \xi^{\beta} \right),
\end{equation}
where $e_d = -1/3$ and $G$ and $I^\prime$ have the same meanings as
before, and
\begin{equation}
M_{\rm spectator\ bremss} =  -{4 e_q G\over m_B }
         C_8(\mu) I_0(\epsilon_B)
          \left( p \cdot q \epsilon \cdot \xi
          + i\epsilon_{\mu\nu\alpha\beta}
          p^{\mu} q^{\nu} \epsilon^{\alpha} \xi^{\beta} \right),
\end{equation}
where $e_q$ is the quark charge for the spectator. We neglected
some small terms in the numerator and let
\begin{equation}
I_0(\epsilon_B) = -i\pi+\ln{1-\epsilon_B\over{\epsilon_B}} .
\end{equation}

The coefficient $C_8$ is
\begin{equation}
C_8(m_W) = -{x\over 8(1-x)^4}\left((x-1) (x^2-5x-2)
                   +6x \ln x \right) = -0.094,
\end{equation}
where the numerical value is again for $x = 4$, and it scales like,
\begin{equation}
C_8(\mu) = \eta^{-14/3\beta_0} \left[ C_8(m_W)
               -{11\over 144}\left(\eta^{8/3\beta_0} -1\right)
               +{35\over 234}\left(\eta^{26/3\beta_0} -1\right)
              \right] .
\end{equation}
As the renormalization scale decreases, the magnitude of $C_8$
decreases, actually passing through zero at $\mu \approx 6$ GeV.

Additionally, there are the annihilation graphs,
Fig.~\ref{diagrams}d.  These can contribute only to
$B^\pm$ decay. To leading order in $(\epsilon_B)^{-1}$ the result
comes from bremsstrahlung off the initial $u$ quark and gives
\begin{equation}
M_{\rm ann} =  {2e_u\over m_B \epsilon_B }  \,
         {m_{K^*} \over m_B}
         \left[ e {G_F \over\sqrt 2} V_{ub} V_{us}^* f_B f_{K^*}
                                                 \right]
          \left( p \cdot q \epsilon \cdot \xi
          + i\epsilon_{\mu\nu\alpha\beta}
          p^{\mu} q^{\nu} \epsilon^{\alpha} \xi^{\beta} \right),
\end{equation}
where we kept $m_{K^*}$ when it appeared as an overall factor and
where $e_u = 2/3$.  The quantity in square brackets differs from
the quantity $G$ used earlier in lacking the strong interaction
factors $C_F \alpha_s / 4\pi$ and in having different CKM factors.
While it is interesting that the decay proceeds at all without
flavor changing neutral currents, the result turns out small. Not
having the gluon exchange is a plus numerically, but the factors
$m_{K^*}/m_B$ and $V_{ub}$ ensure the small result.

\section{Numerical results}

The numerical results should not depend on the renormalization
scale.  However, as we are most familiar with the wave functions
or distribution amplitudes at a typical hadronic scale, say $\mu
\approx 1$ GeV, we should evaluate the other quantities at the
same scale.  There are very big changes in the $C_i$ from their
values at the $W$ mass scale.

For the sake of definiteness we shall use
$$\Lambda_{QCD}=100\ {\rm MeV},$$
$$\bar\Lambda_B =500\ {\rm MeV},$$
$$m_W = 81\ {\rm GeV},$$
$$m_t = 2m_W,$$
$$V_{ts} = -0.045,$$
$$V_{tb}=0.999,$$
$$V_{ub}=0.0045,$$
$$V_{us}=0.22,$$
$$\tau_B = 1.46\ {\rm picoseconds},$$
$$f_B=132\ {\rm MeV},$$  and
$$f_{K^*}= 151\ {\rm MeV}$$
Our convention has $f_\pi = 93$
MeV and the signs of the CKM parameters follow a ``standard''
advocated in~\cite{pdg}. The sign of
$V_{tb}V_{ts}^*/V_{ub}V_{us}^*$ is what is crucial and it does not
depend on conventions.  We also use the asymptotic form for the
$K^*$ distribution amplitude and will mention results with another
form later.  We will express each contribution to the amplitude as
\begin{equation}
M_i = t_i \times {1\over 2} \left(  \epsilon \cdot \xi
          + i(p \cdot q)^{-1} \epsilon_{\mu\nu\alpha\beta}
          p^{\mu} q^{\nu} \epsilon^{\alpha} \xi^{\beta} \right),
\end{equation}
whereupon
\begin{equation}
\Gamma = {1 \over 16 \pi m_B} |t|^2
\end{equation}
for $t$ being the sum of the $t_i$ (and neglecting the $K^*$ mass).

The scale we should use  should be compatible with the scale
that our wave functions and distribution amplitudes are
determined at, and this is turn should be consistent with the
scale of the four-momentum squared of the off-shell gluon.  This
suggests $\mu \approx 0.8$ GeV, which we shall use.
We extrapolate the coefficients according to the formulas given
earlier. Much of the change due to the extrapolation occurs as the
scale changes from $m_W$ down to order $m_B$, at least for the
large terms $C_7$ and $C_2^\prime$.  For the running coupling
we use $\alpha_s = 4\pi / \beta_0 \ln(\mu^2 / \Lambda_{QCD}^2)$
with number of flavors appropriate to the scale.

{}From operators $O_7$ and $O_2^\prime$ we get
\begin{eqnarray}
t_7 &=& -0.95 - 3.56i \nonumber \\
t_{2^\prime} &=& 2.40
\end{eqnarray}
with the contribution from photonic penguins being the sum of
these two. The amplitudes are in units of $10^{-8}$ GeV.
For others we get
\begin{eqnarray}
t_{\rm b\ or\ s\ bremss} &=& 0.08, \nonumber \\
t_{\rm spectator\ bremss} &=& 0.04 - 0.05i ,\nonumber \\
t_{ann} &=& 0.06 .
\end{eqnarray}

The last two are given for the charged $B$.  The photonic
penguins dominate, although the other graphs contribute circa 10\%
corrections to the real parts of the decay amplitude.  The extra
operator that we have considered in the photonic penguin
calculation gives an amplitude with a magnitude that is about $2/3$
as large as the amplitude from the operator normally considered. It
is the luck of the phases that keeps its effect small.  It
increases the rate by not even 10\% since roughly speaking what it
does is just change the sign of the real part of the amplitude.

Specifically, we get a branching fraction
\begin{equation}
Br(B \rightarrow K^* \gamma) = 1.24 \times 10^{-5}
\end{equation}
including just the photonic penguin terms.  Keeping only the
most usual operator $O_7$ would reduce the branching fraction to
$1.13 \times 10^{-5}$.  It seems inconsistent
to include the smaller contributions since they may be smaller
than the errors induced by our approximations upon the big
terms.  However, keeping all terms anyway gives
$1.24 \times 10^{-5}$ (unchanged from above) for the neutral $B$ and
$1.31 \times 10^{-5}$ for the charged $B$.  The relative size of the
neutral and charged $B$ decays should be about right and would be
interesting to observe as more precise data becomes available.

It is possible that the distribution amplitude for the transversely
polarized $K^*$ is narrower than the asymptotic one.  Chernyak,
Zhitnitsky, and Zhitnitsky~\cite{czz} suggested a distribution
amplitude
\begin{equation}
\tilde \phi_{K^*}(y) = 5(y_1y_2)^2,
\end{equation}
albeit this was for the transverse $\rho$.  If we use this
distribution amplitude for the $K^*$, our calculated branching
fractions are roughly halved.

The choice of $\Lambda_{QCD}$ was made consistent with some
of our own earlier work~\cite{cm93,cm94}, but could be varied
(the earlier situation is much less sensitive to the value of this
quantity than the present case will prove to be).  If we let
$\Lambda_{QCD} = 200$ MeV, leaving other parameter choices
untouched, the branching ratio with the asymptotic
distribution amplitude changes to
\begin{equation}
Br(B \rightarrow K^* \gamma) = 3.50 \times 10^{-5}
\end{equation}
including just photonic penguins terms, with commensurate changes in
results keeping all terms or just $O_7$.  These values
are in accord with present experimental data~\cite{cleo93}. It is
also clear that values for other parameters could be varied somewhat
from values that we have used.

\section{Conclusion}

It seems with present knowledge, the actual $B \rightarrow K^*
\gamma$ decay rate is sensitive to parameters of the bound state and
to parameters governing the  evolution of QCD.  Still, a number of
conclusions may be drawn.

Contributions from gluonic penguin and annihilation
diagrams---which contribute to the physical decay but not to
$b\rightarrow s \gamma$---have been calculated here. They change
the decay rate by a few percent and so are not worrisome until the
experiments are considerably more precise.

Also calculated here, and more significant, are effects due to
decay quarks being off-shell. This brings into play another
operator in the effective hamiltonian for
$b\rightarrow s \gamma$, and this new operator
produces an amplitude of noticeable magnitude.  However, its phase is
such that the effect on the decay is under 10\%.

Regarding the future, there is a need to calculate the double
penguin diagrams mentioned in the introduction and illustrated
in Fig.~\ref{diagrams}e,  including the QCD corrections to them.
For now, we can at least note that these diagrams will not
contribute an imaginary part to the amplitude, so that the imaginary
part that is already there puts a lower bound on the total result.
Also, one will wish to eventually study the totality of
$B\rightarrow X_s  \gamma$ since this is closer to the basic
$b\rightarrow s \gamma$ than any individual exclusive channel.
Still, the physical   $B\rightarrow X_s  \gamma$ always
involves spectators and always has a contribution from operators
that only contribute if the $b$ is off-shell.  We have not
calculated for situations other than the lowest ``$X_s$'' and do
not know what size corrections ensue overall, or even if the phase
situation for the off-shell contributions persists.

None-the-less, the opportunity to test the flavor changing neutral
currents induced in the standard model and to search for evidence of
particles or phenomena beyond the standard model makes $B$ decay
into $K^* \gamma$ and into $X_s \gamma$ interesting, and makes
calculations to determine precisely the standard model
contributions to these decays a worthwhile and necessary pursuit.

\bigskip
{\centerline{ACKNOWLEDGEMENTS}}
\medskip

We thank H. J. Lu for many detailed discussions and also S. Dawson,
J. Donoghue, W. Marciano, and A. Szczepaniak for helpful
conversations.  We also thank D. Griegel for his input.  This work
was supported in part by NSF Grant PHY-9306141 and DOE Grant
DOE-FG02-93ER-40762.

\vskip 0.6in

\noindent {\Large {\bf Figure caption}}

{Fig.~\ref{diagrams}.  Diagrams.  The effective vertex, due for
example to a $W$ and $t$ quark loop, is represented as an oval
blob;  (a) shows the photonic penguins, (b) shows gluonic penguins
with bremsstrahlung from the
$b$ or
$s$ quark, (c) shows gluonic penguins with bremsstrahlung from
the spectator quark, (d) shows two of four lowest order
annihilation diagrams that could give charged $B \rightarrow
K^* \gamma$, and (e) shows the ``double penguin,'' once as a blob
and once showing one example of a contribution to the blob.}

\newpage

\begin{figure}
\vglue 7.7in

\hskip 1.15in {\special{picture diagrams scaled 1000}} \hfil

\caption{}

\label{diagrams}
\end{figure}

\end{document}